\documentclass{emulateapj}
\usepackage{apjfonts}
\usepackage{color}

\usepackage{hyperref}
\hypersetup{
    colorlinks=true,
    linkcolor=blue,
    filecolor=magenta,      
    urlcolor=cyan,
}

\newcommand{\code}[1]{\texttt{#1}}
\newcommand{\mesa}{\code{MESA}}
\newcommand{\MESA}{\mesa}
\newcommand{\FLASH}{\code{FLASH}}

\newcommand\beq{\begin{equation}}
\newcommand\eeq{\end{equation}}
\newcommand{\mini}{M_{\rm ini}}
\newcommand{\Msun}{\ensuremath{{\rm M}_\odot}}

\newcommand{\Rsun}{\ensuremath{{\rm R}_{\odot}}}

\shorttitle{Ultra long GRBs from very massive stars}
\shortauthors{Perna, Lazzati \& Cantiello}

\begin{document}

\title{ULTRA LONG GAMMA-RAY BURSTS FROM THE COLLAPSE OF BLUE SUPER GIANT STARS: \\ AN END-TO-END SIMULATION}

\author{Rosalba Perna}
\affiliation{Department of Physics and Astronomy, Stony Brook University, Stony Brook, NY, 11794, USA}

\author{Davide Lazzati}
\affiliation{Department of Physics, Oregon State University, 301
Weniger Hall, Corvallis, OR 97331, USA}

\author{Matteo Cantiello}
\affiliation{Center for Computational Astrophysics, Flatiron Institute, 162 5th Avenue, New York, NY 10010, USA}
\affiliation{Department of Astrophysical Sciences, Princeton University, Princeton, NJ 08544, USA}

\begin{abstract}
  Ultra-long gamma ray bursts (ULGRBs) are a distinct class of GRBs
  characterized by durations of several thousands of seconds, about
  two orders of magnitude longer than those of standard long GRBs
  (LGRBs). The driving engine of these events has not been uncovered
  yet, and ideas range from magnetars, to tidal disruption events, to
  extended massive stars, such as blue super giants (BSG). BSGs, a
  possible endpoint of stellar evolution, are attractive for the
  relatively long free-fall times of their envelopes, allowing
  accretion to power a long-lasting central engine. At the same time,
  their large radial extension poses a challenge to the emergence of a
  jet.  Here we perform an end-to-end simulation aimed at assessing
  the viability of BSGs as ULGRB progenitors.  
  The evolution to core collapse of a BSG star model is calculated
  with the \MESA\ code. We then compute the accretion rate for
  the fraction of envelope material with enough angular momentum to
  circularize and form an accretion disk, and input the corresponding
  power into a jet which we evolve through the star envelope with the
  \FLASH\ code.  Our simulation shows that the jet can emerge, and the
  resulting light curves resemble those observed in ULGRBs, with
  durations $T_{90}$ ranging from $\approx\,4000$~s to $\approx\,10^4$~s
  depending on the viewing angle.
\end{abstract}


\section{Introduction}

The class of long Gamma-Ray Bursts (LGRBs), lasting longer than about
2~seconds and with average duration of several tens of seconds, has
been unambiguously associated with the core collapse of massive stars.
In particular, the direct association of some LGRBs with Type Ic
supernovae (SNe; Stanek et al. 2003; Hjorth et al. 2003), implies that
these LGRBs are connected to the death of Wolf-Rayet (WR) stars. By
the time of the iron core collapse, these stars have lost their
hydrogen-rich envelopes, and hence they are significantly more compact
than common Type~II supernova progenitors. In the standard collapsar
scenario (MacFadyen \& Woosley 1999), as the core collapses into a
black hole (BH), material
from the outer layers rains back and forms an accretion disk, if
endowed with enough angular momentum to circularize outside of the
last stable orbit.  Rapid accretion of this disk onto the BH powers a
powerful engine. This energy, tapped either via neutrinos (Quian \&
Woosley 1996; Popham, Woosley \& Fryer 1999; Kohri \& Mineshige 2002;
Lee, Ramirez-Ruiz \& Page 2005) or the Blandford-Znajek effect
(Blandford \& Znajek 1977; see also De Villiers, Hawley \& Krolik
2003; Tchekhovskoy, Narayan \& McKinney 2011), launches a relativistic
jet, which has to punch out of the thick envelope before being able to
dissipate and produce gamma-rays. The small size of a WR star
facilitates the emergence of the jet (e.g. Matzner 2003).  This,
combined with the fact that the free-fall time from the outer layers
of these stars is several tens of seconds, provides a further
theoretical support to the association between long GRBs and WR stars.

In the last several years, a number of GRB events with considerable
longer duration has been detected.  These bursts, named Ultra Long
GRBs (ULGRBs), are characterized by a gamma-ray emission that lasts
for several thousands of seconds. When considering the $\gamma$-ray
component alone, their duration makes them statistically distinct from
traditional LGRBs (Boer et al. 2015; Levan 2015).  However, if one
measures the burst duration also including the X-ray component, which
displays plateaus and flares over timescales of several thousands of
seconds in many LGRBs, then the evidence for a separate GRB class
becomes less clear (Zhang et al. 2014).  Theoretically, plateaus and
flares have been explained with a variety of models, including
magnetars (Rowlinson et al. 2014) and specific progenitor envelope
structures for plateaus (Kumar, Narayan \& Johnson 2008), and
properties of the accretion disk for flares (Perna, Armitage \& Zhang
2006; Proga \& Zhang 2006).  In the case of the ULGRBs, the models
which have been proposed essentially fall within three classes: tidal
disruption (TD) of a white dwarf (WD) by a black hole (Gendre et
al. 2013; Levan et al. 2014; MacLeod et al. 2014; Ioka, Hotokezaka \&
Piran 2016); a newly born magnetar (Greiner et al. 2015; Gompertz \&
Fruchter 2017); or fallback accretion from the envelope of an extended
progenitor star, such as a blue supergiant (BSG; Quataert \& Kasen
2012; Wu et al. 2013; Nakauchi et al 2013; Liu et al. 2018). For the
latter scenario, Suwa \& Ioka (2011) have suggested that jet breakout
is possible in supergiant stars if the envelope accretion and
resulting central engine are sufficiently long-lived. Simulations by
Nagakura et al. (2012) support this scenario, at least for some
moderately extended stellar envelopes from Pop~III massive stars.
However, these calculations assumed that all the accreted
mass powers a jet, without considering whether the accreting gas in
the pre-SN envelope has enough angular momentum to
circularize and form a disk.

Generally speaking, for a successful accretion-powered GRB from a
massive star, the star at core collapse has to possess material with
specific angular momentum larger than the minimum required to
circularize at the innermost stable orbit (Woosley 1993; Yoon \&
Langer 2005; Yoon et al. 2006; Woosley \& Heger 2006). The location of
this material within the star sets its free-fall time, and ultimately
influences the duration of the LGRB, if the jet is able to emerge
from the envelope.

In this paper we assess whether the required conditions
described above can be realized during the collapse of a massive,
radially extended progenitor star. This is achieved via an end-to-end
simulation of a relativistic jet propagating through a BSG
progenitor. We begin by simulating the evolution and collapse of a
massive, rotating BSG star with the \MESA\ code (Paxton et al. 2011;
\S2), and numerically compute the fallback rate of the
rotating, collapsing envelope by following the particle trajectories
(\S3).  The fraction of fallback mass which has enough angular
momentum to circularize in an accretion disk provides the power to
launch a jet.  We simulate  the jet evolution through the stellar
envelope with the code \FLASH\ (Fryxell et al. 2000), and compute the
resulting light curves as a function of the viewing angle with respect
to the jet axis (\S4).  Our results, which are discussed in \S5, show that
ULGRBs can indeed be produced via this scenario.

\begin{figure*}
\begin{minipage}[t]{0.5\textwidth}
\centering
\includegraphics[width=9.4cm]{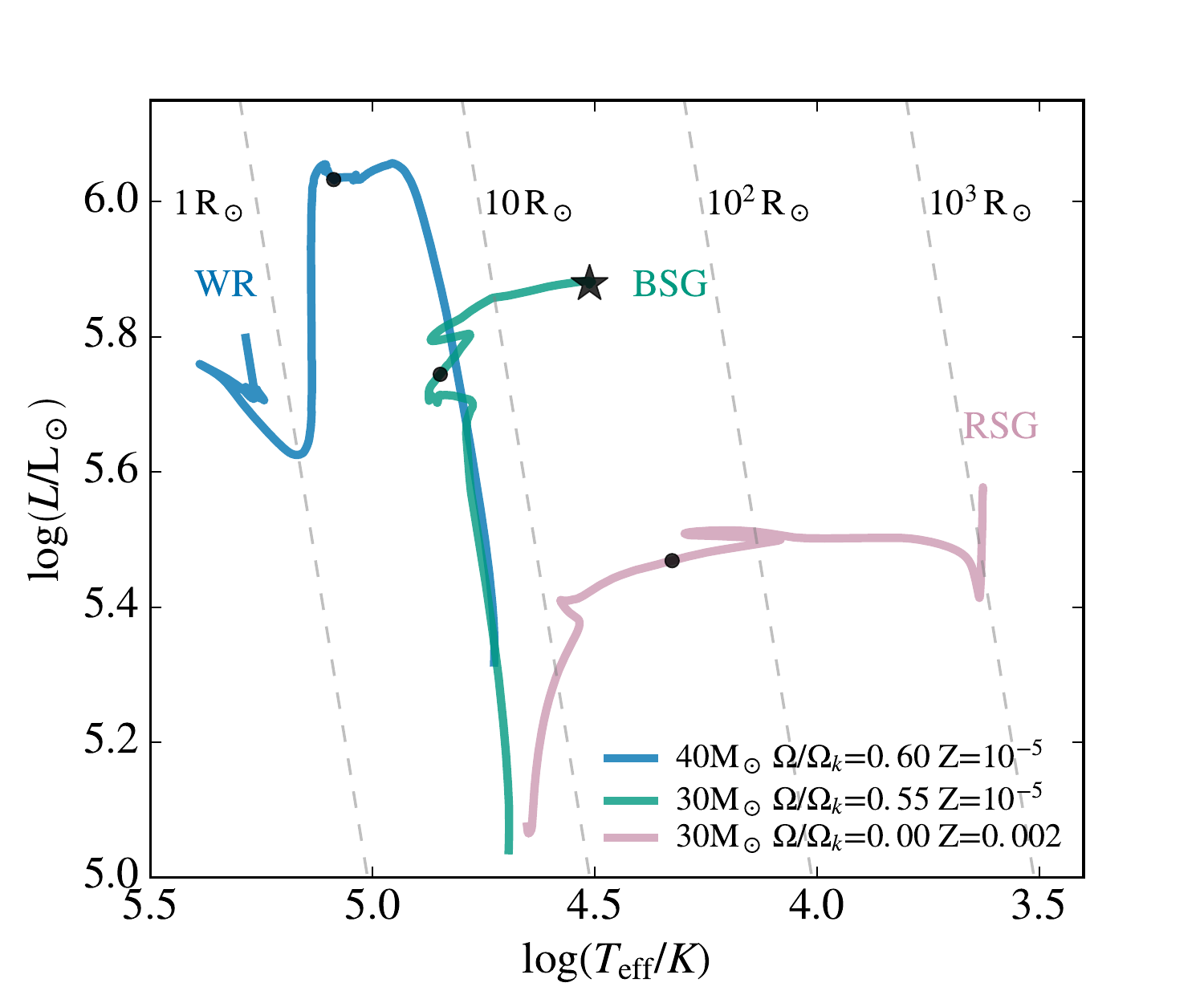}
\end{minipage}
\begin{minipage}[t]{0.5\textwidth}
\centering
\includegraphics[width=9.5cm]{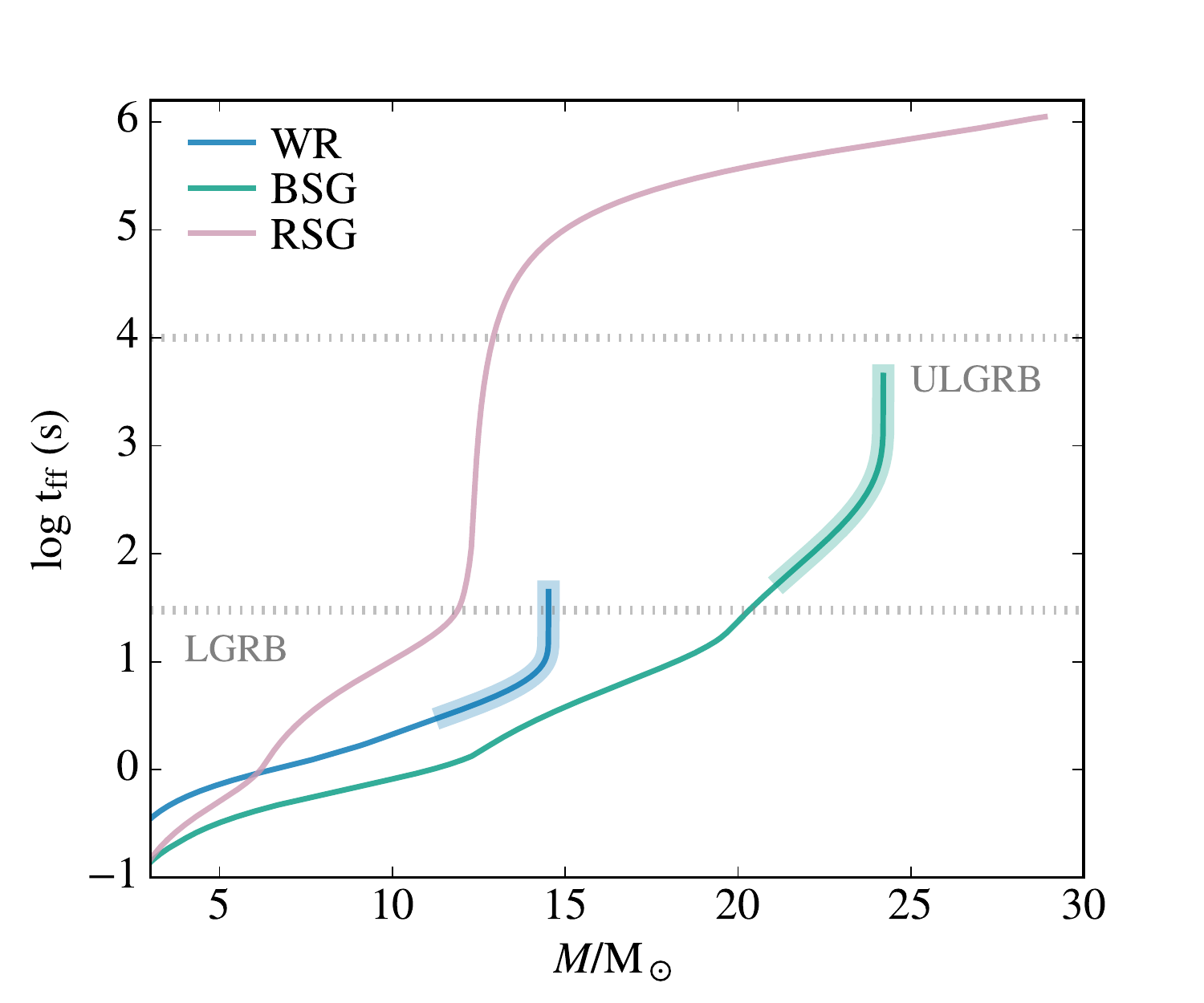}
\end{minipage}
\caption{{\em Left}: HR-Diagram showing the evolution of three massive
  stars {from the zero age main sequence to core collapse},
  calculated with \MESA. A non-rotating massive star with $\mini =
  30~\Msun$ and metallicity Z=0.002 evolves as an RSG and is
  characterized by a large stellar radius. This is a Type II SN
  progenitor model. A star with $\mini = 40~\Msun$, Z=$10^{-5}$ and
  initial rotation rate $\Omega/\Omega_{k}=0.6$ evolves chemically
  homogeneously as a result of rotational mixing. It stays compact and
  becomes a WR. In the literature, similar models are discussed as
  possible progenitors of LGRBs.  Our BSG progenitor star has $\mini =
  30~\Msun$, Z=$10^{-5}$ and initial rotation rate
  $\Omega/\Omega_{k}=0.55$. Unlike our WR star, it does not evolve
  fully chemically homogenously, although its envelope experiences a
  large degree of mixing. This keeps the star blue while preserving an
  envelope of a few tens $\Rsun$. {The star symbol shows the
    location of the BSG model used as input for the calculations
    described in Sec.~\ref{accretion} and Sec.~\ref{GRB}.}  Lines of
  constant radii (dashed lines) are added for reference. {The dot
    symbols show the location where He burning starts contributing
    substantially (90\% of the stellar luminosity).} {\em Right:}
  Free-fall timescale as function of mass coordinate for the three
  models shown in the left panel. The free-fall timescale calculation
  is approximate, but shows how the BSG model accretion timescale is
  compatible with ULGRB durations, while for the WR model it is
  consistent with LGRB durations (see horizontal dotted lines). The
  highlighted part of the curves shows material for which the specific
  angular momentum is larger than the specific angular momentum of the
  last stable orbit around a maximally spinning black hole, and is
  therefore expected to form an accretion disk.}
\label{HRD}
\end{figure*}

\section{Stellar Model}

In stellar evolution calculations, models for supergiant stars are
often found either on a red or a blue solution. Intermediate solutions
are thermally unstable (Woosley, Heger \& Weaver 2002). In the
presence of extreme mass loss, degree of mixing or binary
interactions, some massive star calculations can also move towards
very high temperatures, ending their evolution as compact WR models.
The dichotomy between the LGRB and ULGRB duration can then be
interpreted as resulting from the different activity timescale of a
central engine that is fed by stellar progenitor envelopes of
different radial extension (see Fig.\ref{HRD}).  Note that for the
core collapse of a Red Super Giant (RSG) progenitor, where envelope
material can accrete onto a central black hole due to some fallback
or because the SN explosion failed, the extreme size of the envelope
imposes a very long ($\sim 10^{5 - 6}$s) accretion
timescale. Therefore, even if the material has enough angular momentum
to circularize around the newly formed central object, this can only
result in a very small accretion rate. Such accretion rate may not be
able to power a jet with enough luminosity to burrow through the
prodigious envelope of a RSG, though only dedicated end-to-end
simulations can give a definite answer

For single massive star models, the location on the
Hertzsprung-Russell Diagram (HRD) at the end of their evolution
depends sensitively on the treatment of mass loss, envelope
convection, semiconvection, overshooting and rotational mixing
(e.g. Woosley, Heger \& Weaver 2002; Maeder \& Meynet 2012; Langer
2012; Jiang et al. 2015).  The physics of these processes 
 is not fully understood, and their 1D implementations
should be considered highly uncertain (e.g. Smith 2013; Cantiello et
al. 2014; Jiang et al. 2015).  To complicate the picture, the majority
of massive stars is found in binary or multiple systems (Sana et
al. 2012), with a large fraction of massive stars interacting with one
companion during their lifetime (De Mink et al. 2014). By changing the
size, composition and rotational properties of the stellar envelope,
binary interactions can drastically affect the position of a star in
the HRD (Langer 2012). In this paper we consider a rapidly-rotating,
low-metallicity, single massive star model that ends its life as a
blue supergiant with properties that make it attractive as a possible
ULGRB progenitor.  Given the aforementioned uncertainties, we make no
claim that our particular choices of physics should be preferred
and/or that this evolutionary channel is the most likely to produce
ULGRBs. We just show one possible realization of such evolutionary
pathway.

We use Modules for Experiments in Stellar Evolution (\MESA, Paxton et
al. 2011, 2013, 2015, 2018; release 9793) to create a BSG
progenitor model.  In our calculations the convective boundaries are
determined using the Ledoux criterion and convection is included in
the mixing length (MLT) approximation with $\alpha_{\rm MLT}=1.82$.  We adopt
semiconvection with an efficiency $\alpha_{\rm sc}$ = 0.1 (Langer et
al. 1983, 1985).  The effect of stellar winds is included according to
the mass loss scheme described in Glebbeek et al. (2009; Dutch wind
scheme) with an efficiency parameter $\eta_{\rm wind}$ = 0.8.  We include the
physics of internal angular momentum transport accounting for shear
instabilities, Eddington-Sweet circulation and magnetic torques. We
also account for rotationally enhanced mass loss (Paxton et al. 2013,
Perna et al. 2014). {Models are evolved from the zero age main sequence to iron core formation.}

Our BSG candidate has initial mass $\mini = 30 \Msun$ and metallicity
Z=$10^{-5}$, corresponding to 1/1400 Z$_\odot$.  The calculation
starts on the zero age main sequence with 55\% of Keplerian velocity,
corresponding to an equatorial surface velocity of 530~km~s$^{-1}$.
Mostly due to the efficient chemical mixing associated with the
Eddington-Sweet circulation in rapidly rotating, massive stars (Maeder
1987, Yoon \& Langer 2005, Woosley \& Heger 2006), the star evolves
almost fully mixed during the main sequence. However, unlike the more
massive ($\mini = 40 \Msun$) and rapidly rotating model in
Fig.\ref{HRD}, which stays quasi-chemically homogenous well after its
main sequence becoming a compact WR star, this model manages to
develop a larger compositional gradient in its envelope. This
compositional gradient allows the model to evolve past its main
sequence maintaining a mild core-envelope structure. The large degree
of main sequence mixing, however, results in a stellar envelope that
is very helium rich. Due to the resulting high mean molecular weight
in the envelope, and the reduced line-driven winds at such
low-metallicities, the stellar model ends its life as a blue
supergiant and with a large amount of angular momentum in its He-rich
envelope.  The WR and RSG progenitors shown for reference in
Fig.~\ref{HRD} were calculated with the same physics prescriptions as
for the BSG model, but different values of initial parameters. {
  The bifurcation in the HRD corresponding to rapidly rotating tracks
  with slightly different values of initial rotation (our BSG and WR
  models) has been discussed in the literature (see e.g. Maeder 1987),
  and is due to the sensitivity of rotational mixing to the
  development of compositional gradients.}

A comparison between estimates of the free-fall timescale as a function
of the enclosed mass (see e.g. Woosley \& Heger 2012), \beq t_{\rm
  ff}\,=\,\frac{1}{\sqrt{24\pi G \bar{\rho}}}\,,
\label{eq:tff}
\eeq
clearly shows the difference between a WR, a BSG
and a RSG star (cfr. right panel of Fig.~\ref{HRD}). In the equation
above, $ \bar{\rho}$ is the average enclosed stellar density for a
given stellar model at core collapse. In the figure we highlight
regions of the stellar models (thicker line) corresponding to material
with enough angular momentum to circularize in a disk outside of the
last stable orbit of a black hole (assumed maximally spinning
here). These regions can power a central engine, and their estimated
free-fall timescales can help determining what type of energetic
transients they might produce.  The difference between the typical
timescale of tens of seconds of a WR star, and the tens of thousands
of seconds of a BSG star, is clearly evident. For the non-rotating RSG
model no regions can circularize, and no central engine is expected to
operate.

\section{Accretion rate and jet luminosity}\label{accretion}

The calculation of the rate $\dot{m}_{\rm fb}$ with which  matter
from the collapsing envelope falls back and lands on the
equatorial plane is performed using the formalism by Kumar et
al. (2008).  At $t=0$, the gas is distributed
according to its core-collapse profile, with a velocity field entirely in the
$\hat{\phi}$ direction.  As the star explodes, the fate of the
envelope is sensitively dependent on the explosion energy.  For
energies $\lesssim 10^{51}$~ergs, and stars up to $\sim 40 \Msun$,
the entire envelope falls back (Perna et al. 2014).  For simplicity
here, we assume that subsequent to the star explosion particles
follow a free-fall trajectory, which
depends on their initial angular velocity $\Omega(r,\theta)$ and
location $(r,\theta,\phi)$ in the pre-SN profile.  This
assumption implies that we are neglecting pressure forces and 
any modification to the dynamics due to the SN explosion.
By neglecting these effects, the results of our calculations lead to the most
conservative estimate for the duration of activity of the engine
(shortest fallback timescales).

\begin{figure}
\includegraphics[scale=0.61]{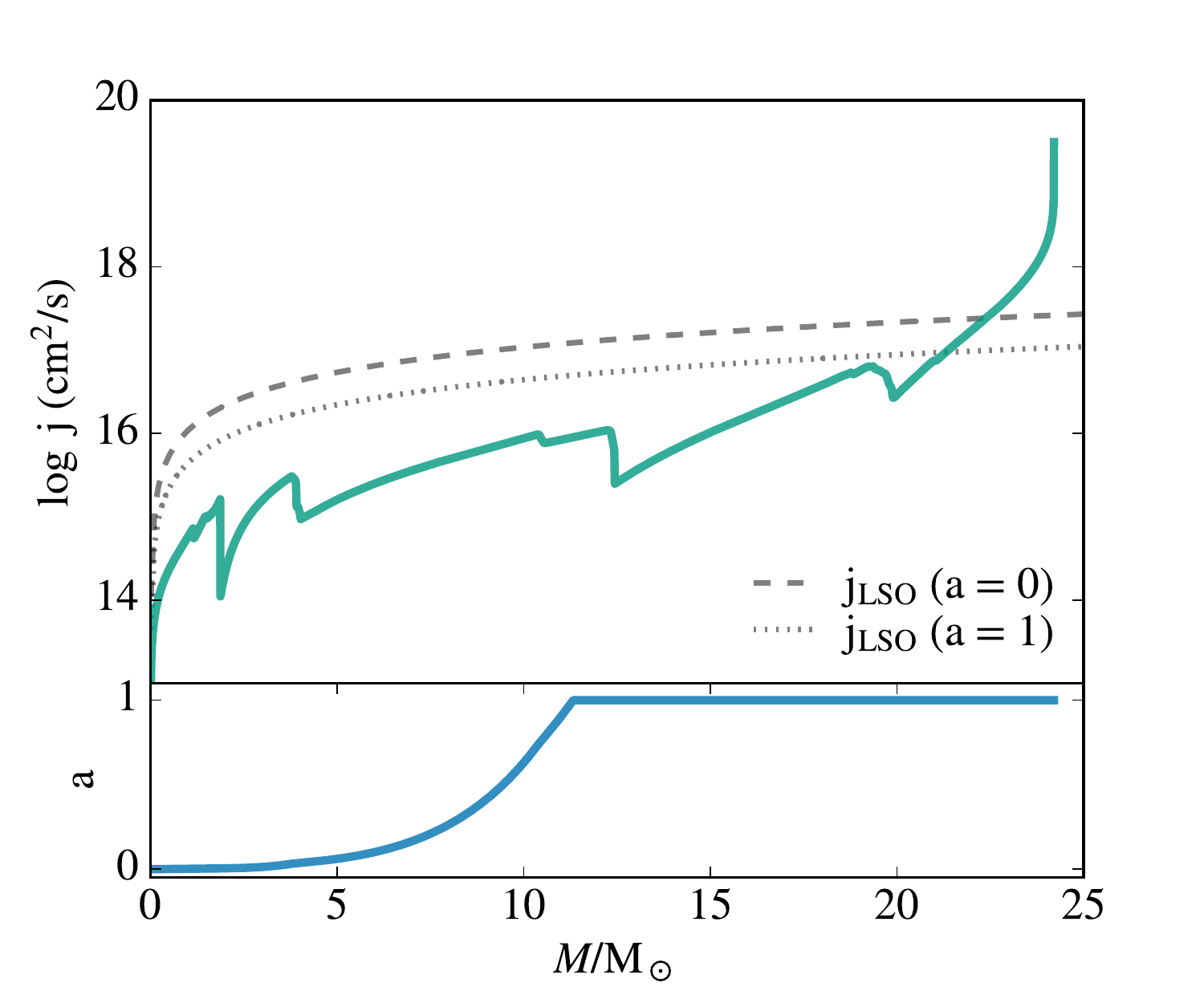}
\caption{{\em Upper panel}: The angular momentum distribution at core
  collapse for the BSG star model considered here. The dotted and dashed
  lines show the specific angular momentum distribution of a particle
  corotating a BH at the last stable orbit.  The dashed line is for
  the case of a BH with zero spin, while the dotted line for a BH with
  maximum spin $a=1$.  The mass of the BH is equal to the enclosed
  mass.  {\em Bottom panel}: BH spin parameter as a function of enclosed 
  mass.  }
\vspace{0.2in}
\label{fig:profile}
\end{figure}

The trajectory of a particle initially positioned at the location
$(r,\theta,\phi)$ with angular velocity $\Omega$ is characterized by
eccentricity $e$,  given by
the expression 
\beq 
e\;=\;1\;-\;\frac{\Omega^2}{\Omega^2_{\rm
    K}}\sin^2\theta\;,
\label{eq:a-e}
\eeq 
where $\Omega_{\rm K}=(GM_r r^{-3})^{1/2}$ is the Keplerian
angular velocity of a particle at location $r$ enclosing a mass $M_r$.  
The particle reaches
the equatorial plane after a time (Kumar et al. 2008) \beq t_{\rm eq}
(r,\theta) = \Omega_{\rm K}^{-1}[ \cos^{-1} (-e) + e(1-e^2)^{1/2}] \;
(1+e)^{-3/2} + t_s(r)\;,
\label{eq:teq}
\eeq where the timescale $t_s (r)\approx \Omega_{K}^{-1}$ corresponds
to the sound travel time from the center to the position $r$. Within
the context of Eq.~(\ref{eq:teq}), it can be interpreted as the time
it takes for information on the loss of pressure after
collapse to propagate to position $r$, after which gas from that
location can begin to collapse.

The  amount of  envelope mass that falls back on the equatorial 
plane between time $t$ and $t+dt$ is  given by
\begin{eqnarray}
dM_{\rm } &=& \dot{m}_{\rm fb} (t) dt = 2\pi \int_{r(t)}^{r(t+dt)} dr\; r^2\\
&\times & \int_{\theta(t)}^{\theta(t+dt)} d\theta\; \sin\theta \;\rho(r,\theta)\,
\delta[t-t_{\rm eq}(r,\theta)]\;.
\label{eq:mdot}
\end{eqnarray}

Once particles reach the equatorial plane, they can circularize only if their specific
angular momentum is equal to or larger than
\begin{equation} j(R) = \frac{\sqrt{GMR} \left[R^2 -                                                    
    2(a/c)\sqrt{GMR/c^2} +(a/c)^2\right]} {R\left[ R^2-3GMR/c^2 +                                       
    2(a/c)\sqrt{GMR/c^2}\right]^{1/2}}\;                                                              
\label{eq:jGR}                                                                                          
\end{equation}
at the last stable orbit, where $j$ represents the specific
angular momentum of a particle on a co-rotating orbit of a black hole
of mass $M$ and angular momentum $J=aM$.  Here the coordinate $R$ 
describes the radial distance in the disk plane.  

\begin{figure}
\includegraphics[scale=0.61]{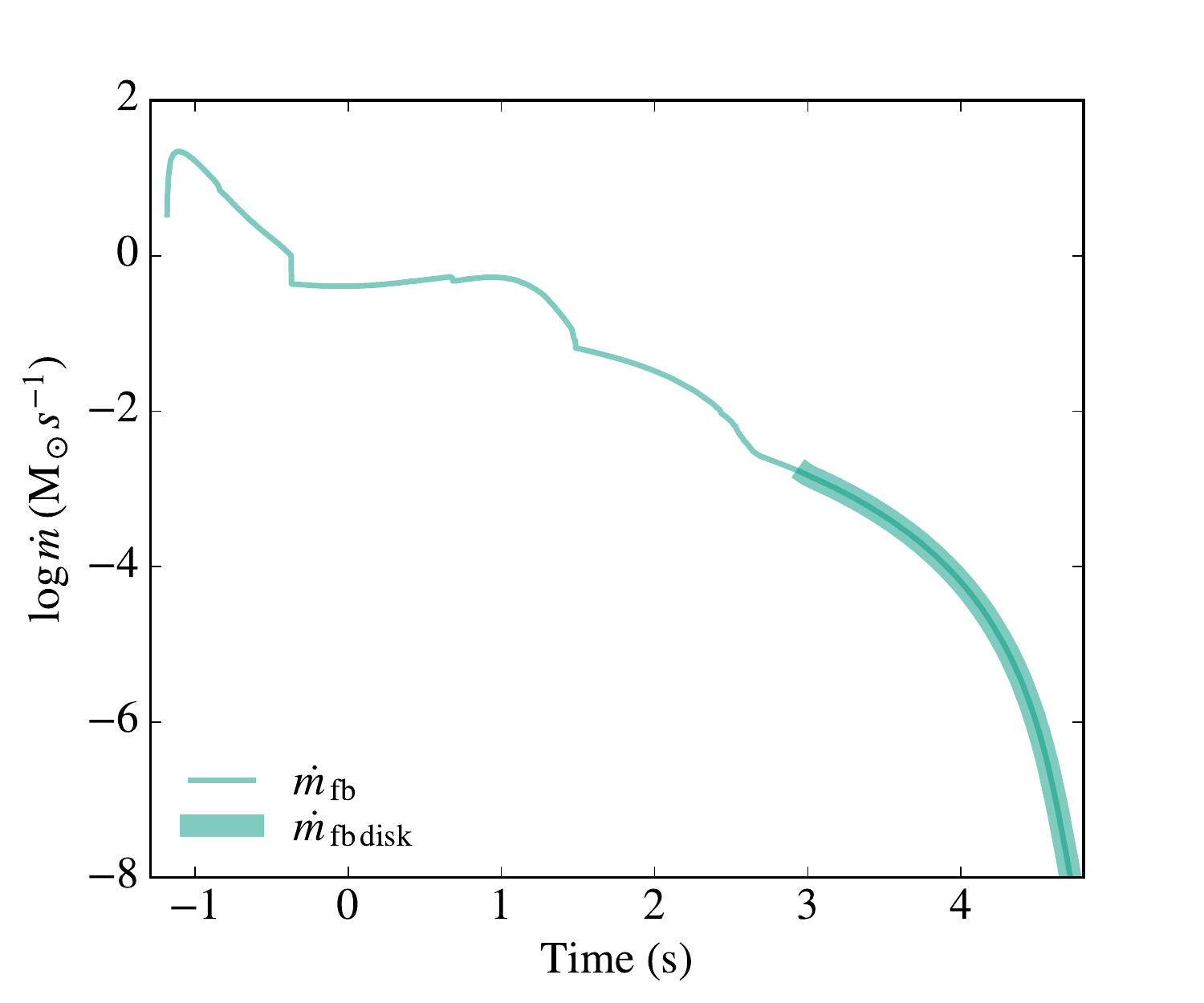}
\caption{Fallback mass rate $\dot{m}_{\rm fb}$ of our BSG model as
  function of time after core collapse. The highlighed portion of the
  curve shows the component $\dot{m}_{\rm fb, disk}$ having enough
  specific angular momentum to circularize in a disk. }
\label{fig:mdot}
\end{figure}

For the specific case that we are studying here, the angular momentum
distribution of the pre-SN profile is shown in
Fig.~\ref{fig:profile}. As a reference, we also display the specific
angular momentum distribution of a particle corotating a BH at the
last stable orbit for both a non-rotating and a maximally-rotating BH.
Inspection of the pre-SN profile shows that only the outermost layers
of the envelope have sufficient angular momentum to circularize.
Hence, while computing the rate of fallback mass onto the equatorial
plane, we further track the component $\dot{m}_{\rm {fb, disk}}$ of
this rate that has sufficient angular momentum to form a disk.  This
is the relevant quantity for our simulations.  We display
$\dot{m}_{\rm{fb,disk}}$ in Fig.~\ref{fig:mdot}, together with the
total fallback accretion rate $\dot{m}_{\rm {fb}}$.  It is evident that
the onset of an accretion-powered engine
only happens at $\sim 800-900$~s after the explosion.

Given a certain $\dot{m}_{\rm{fb,disk}}$, the fraction of matter which
accretes onto the BH is likely not a fixed fraction of
$\dot{m}_{\rm{fb,disk}}$. At high accretion rates and small radii,
cooling is dominated by neutrino emission, and the flow can be
described by a neutrino-dominated accretion model, or NDAF (Popham et
al. 1999; Di Matteo et al. 2002; Kohri \& Mineshige 2002; Janiuk et
al. 2004, 2007; Lee et al. 2004, 2005; Kohri et al. 2005). Under these
conditions all the matter accretes onto the BH, and hence we can
assume $\dot{m}_{\rm acc}=\dot{m}_{\rm fb, disk}$.  Note that this
condition also relies on the fact that the viscous timescale of the
disk is much shorter than the fallback timescale, and hence matter
flows through the disk at virtually the same rate at which it falls
back onto the disk and replenishes it. However, while the condition
$\dot{m}_{\rm acc}=\dot{m}_{\rm fb, disk}$ safely holds in the NDAF
regime, at lower accretion rates, the flow is expected to switch to
an advection-dominated accretion flow (ADAF, Narayan \& Yi 1994;
1995), with only a fraction of the gas reaching the BH, and with some
mass lost to winds (Stone et al. 1999). Given the uncertainties
associated with the transition from NDAFs to ADAFs, and the lack of a
specific functional form for the amount of mass lost to winds as a
function of radius, in the following we will assume that the condition
$\dot{m}_{\rm acc}=\dot{m}_{\rm fb, disk}$ always holds (where by
$\dot{m}_{\rm acc}$ we hence intend the mass that reaches the BH).  

{Another point to note, which was made by Chen \& Beloborodov
  (2007), is that, during the NDAF phase of the accretion disk, an
  outflow is likely to be launched only above a critical 'ignition'
  accretion rate, $\dot{M}_{\rm ign}=K_{\rm ign}(\alpha_{\rm
    SS}/0.1)^{5/3}$ (where $\alpha_{\rm SS}$ is the viscosity parameter,
  Shakura \& Sunyaev 1977), above which the neutrino flux rises
  dramatically.  The value of the constant $K_{\rm ign}$ was found to
  vary between 0.071~\Msun for a spin parameter $a=0$, and 0.021~\Msun
  for $a=0.95$ (Chen \& Beloborodov 2007). Inspection of
  Fig.~\ref{fig:mdot} shows that, for our BSG star, such high rates are
  achieved during a time at which the material that falls back has not
  enough angular momentum to circularize. The accretion rates which
  are expected to power the jet are lower than $\dot{M}_{\rm
    ign}$. Hence an implication of our model is that the jet-launch
  mechanism must be of magnetic origin (Blandford \& Znajek 1977; De
  Villiers et al.  2003; Tchekhovskoy et al. 2011).} { MHD
jets from collapsing stars are expected to be stable to kink instabilities
(Margalit et al. 2017).}

Finally, the jet luminosity can be written as 
\beq
L_{\rm jet}\;=\; \eta\,\dot{m}_{\rm acc}\,c^2\;,
\label{eq:ljet}
\eeq where $\eta$ is an efficiency factor which parametrizes the
fraction of mass converted into energy to power the jet.
General-relativistic magneto-hydrodynamic simulations of accretion
disks show that the efficiency in converting torus mass and BH spin
into jet energy varies between a few percent up to $\approx 100$\% for
maximally spinning BHs (De Villiers et al. 2005; Tchekhovskoy et
al. 2011; McKinney et al. 2012; Fragile et al. 2012). The efficiency
depends sensitively on the BH spin, the disk thickness, and the
magnetic flux. Here we adopt a low value $\eta=0.01$.  Since the
ability of the jet to bore through the stellar envelope depends sensitively
on its luminosity, our assumption is very conservative.  Larger
values would result in more powerful jets.

{Before continuing with the discussion of the jet
  propagation within the pre-collapse envelope, we need to emphasize
  that, since our simulations do not treat the actual SN explosion, we
  assume that the envelope in which the jet propagates remains fixed
  to its pre-SN distribution. In reality, as the star explodes and the
  jet activity pushes material along the jet axis, the mass
  distribution will change. This effect was discussed by Gilkis, Soker
  \& Papish (2016) in the context of super-energetic SNe (see also Lindner
et al. 2010 in the collapsar scenario), bit it
would similarly apply to our ULGRB scenario. They found that
material previously in hydrostatic equilibrium will start moving outward due to
the reduced gravity, further extending the duration of the accretion phase. }

\begin{figure}
\hspace{-0.3in}
\includegraphics[scale=0.35]{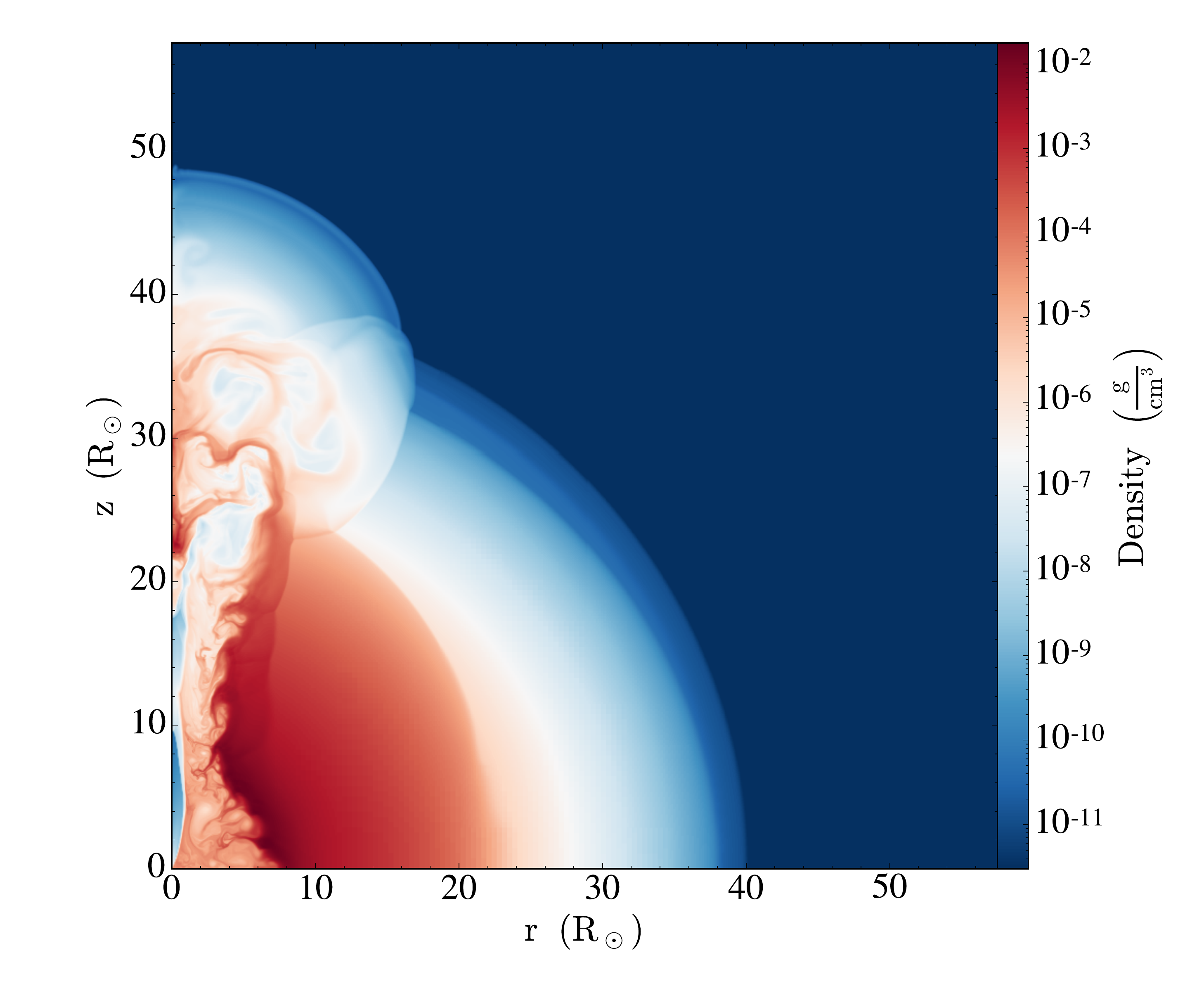}
\caption{Snapshot showing the density structure 
400~s after jet launch in our \FLASH\ simulation.
The jet breaking out of the BSG envelope structure is clearly visible.}
\label{fig:simula}
\end{figure}

\section{Jet evolution and GRB luminosity}\label{GRB}

The simulation of jet propagation in the pre-SN envelope is performed
with the adaptive mesh refinement relativistic hydrodynamic (AMR- RHD)
code \FLASH\ (Fryxell et al. 2000), as modified in Morsony et al.
(2007; see Figure~\ref{fig:simula} for a pseudocolor rendering of the
density at $t=400$~s after the jet launch).  The jet starts at
$t_0=989$~s as a boundary condition with time-dependent luminosity
given by Eq.~(\ref{eq:ljet}), where the accretion rate is provided by
the fallback mass component able to circularize in a disk prior to
accreting ($\dot{m}_{\rm fb, disk}$; cfr. highlighted curve in
Fig.~\ref{fig:mdot}). At earlier times, the accreted mass falls
directly into the BH and a jet is not launched. The jet is injected
with an opening angle $\theta_{j}=16^{\circ}$, compatible with the
{ range of values inferred via different analyses 
for} GRB~111209A ($\theta_j\approx 9^\circ-11^\circ$, Kann et al. 2017, and
$\theta_j=23^\circ$ Stratta et
al. 2013) and the constraint for GRB~121027A ($\theta_j>10^\circ$;
Levan et al. 2014). It should be noted, however, that the injection
opening angle can be very different from the opening angle of the cone
within which the outflow propagates after breakout off the stellar
envelope (e.g., Morsony et al. 2007; see also the discussion below).
The jet is injected with an initial Lorentz factor $\Gamma_0=5$, and
with internal energy that enables acceleration to an asymptotic
Lorentz factor $\Gamma=300$, values that are commonly adopted in long
GRB simulations (e.g., Morsony et al. 2007). The computational grid
extends in the radial direction from $r_0=10^{10}$~cm to $r_{\rm
  max}=4\times 10^{12}$~cm. The maximum resolution, at the base of the
jet, is $\sim5\times10^7$~cm. A polytropic equation of state with
adiabatic index $\hat\gamma=4/3$ is adopted throughout the domain. The
numerical simulation was run for 1803~seconds, after which the
jet-star interaction wanes and the jet luminosity becomes a constant
fraction of the luminosity injected at the base.

The light curves as a function of the viewing angle are computed
following Morsony et al. (2010) by assuming that a constant fraction
of the kinetic luminosity carried by the jet is radiated:
\begin{equation}
L(t)=c\epsilon\int_{\Sigma_R}\left[\left(4p+\rho c^2\right)\Gamma^2-\rho c^2\Gamma\right]\delta^2
d\sigma\,,
\end{equation}
where $\epsilon$ is the radiative efficiency, $\Sigma_R$ is a
spherical surface of radius $R$ centered on the GRB engine, $p$ and
$\rho$ are the pressure and comoving density of the jet, respectively,
and $\delta= [\Gamma(1 - \beta \cos \theta)]^{-1}$ is the Doppler
factor. We calculate the light curves at a radius $R=3\times10^{12}$ cm,
when the jet has left the stellar surface, and at a safe distance form
the outer boundary of the simulation domain. We assume a radiative
efficiency $\epsilon=0.5$ (Zhang et al. 2007; see also discussion on
efficiencies in Giacomazzo et al. 2013), and consider only emission
from material moving with a Lorentz factor $\Gamma\ge10$.
Material with lower Lorentz factor does not substantially contribute to the
radiated luminosity.

\begin{figure}
\includegraphics[scale=0.62]{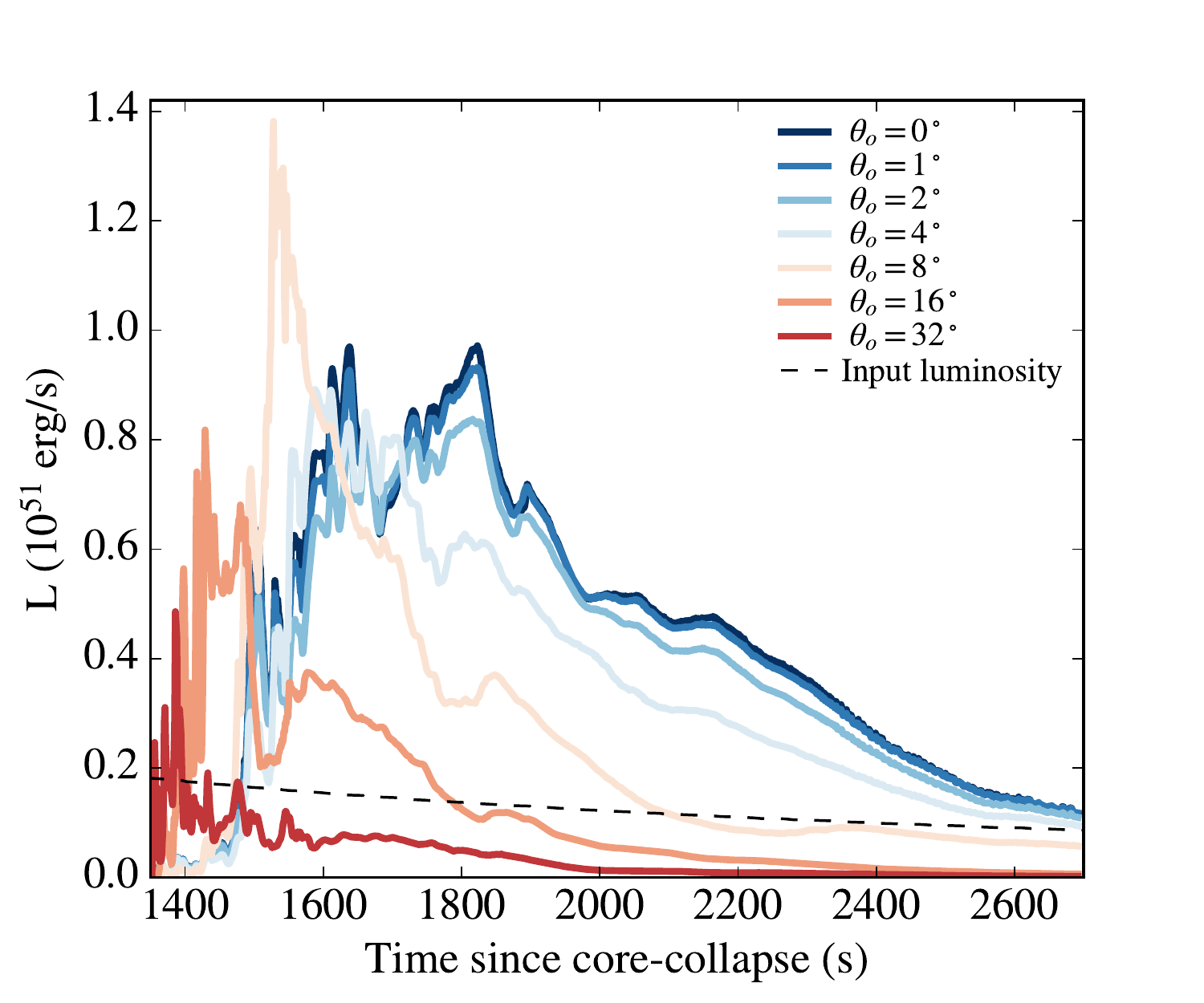}
\caption{Isotropic equivalent bolometric luminosity of the prompt
  emission, as a function of time from core collapse (the jet is
  launched at $t=989$~s in these units and jet breakout occurs
around $1389$~s), and for a range of viewing
  angles. At early times, the luminosity is higher than the input one
  $L_{\rm jet}$ (also shown as isotropic equivalent) since the jet is
  squeezed by the star pressure into a smaller angle than the one at
  injection, and because of the sudden release of the cocoon energy,
  which was trapped within the star up to jet breakout.}
\label{fig:lc}
\end{figure}

\begin{figure}
\includegraphics[scale=0.58]{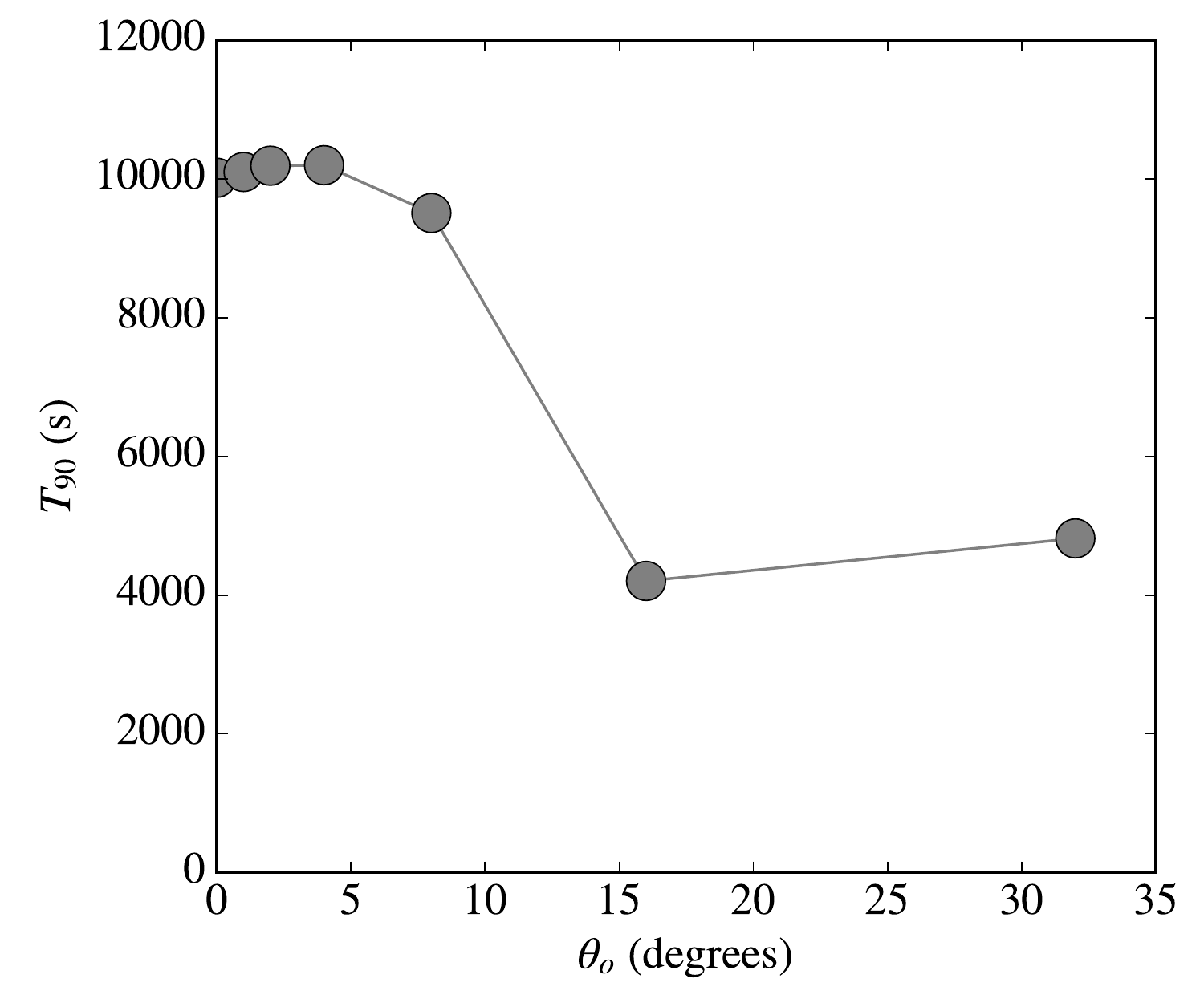}
\caption{$T_{90}$ duration of the bursts from our simulation as a
  function of the observer angle.  Note that the duration becomes
  shorter at viewing angles larger than the opening angle at
  injection. See text for a discussion.}
\label{fig:t90}
\end{figure}

Fig.~\ref{fig:lc} shows the isotropic equivalent bolometric light
curves for a range of viewing angles.  Luminosities of the order of
$10^{51}$~erg/s are seen for observers lying within a few tens of
degrees from the jet axis.  The bright part of the light curve lasts
for approximately 1000~s, followed by a smooth decline that tracks the
decay of the engine luminosity.  We compute the observed duration of
our predicted bursts ($T_{90}$) by integrating the light curves and
finding the times at which 5\% and 95\% of the fluence has been
radiated.  We stop our numerical simulation at 1803~seconds, since the
light curve is well represented by a re-scaled version of the input
luminosity at longer times. The resulting duration ($T_{90}$) is of
the order of $10^{4}$~s, in good agreement with observed ULGRBs (see
Figure~\ref{fig:t90}). At large viewing angles the duration is
shorter. Qualitatively, one can understand this behavior in the
following way. As the jet propagates through the stellar progenitor it
produces a cocoon (Ramirez-Ruiz et al. 2002; Lazzati \& Begelman 2005)
that recollimates the jet into an angle that is narrower than the
injection value $\theta_j=16^\circ$. When the jet breaks out of the
stellar envelope, the outflow is composed of a fast core of a few
degrees opening angle surrounded by slower wings extending to large
angles $\theta\sim45^\circ$. The wings are due to the shearing of the
jet sides on the cocoon material and to the cocoon material itself
that has high pressure and, once released on the stellar surface, is
free to accelerate quasi-isotropically. As time progresses, the
confining pressure of the stellar material decreases and the jet
widens to its injection value.

As seen in Figure~\ref{fig:lc}, light curves for observers within
$\theta_o~\lesssim~5^\circ$ are characterized by high luminosity,
sustained for at least 1000~s, followed by a decline similar to the
luminosity of the central engine. Note that, in the early phase, their
isotropic-equivalent luminosity is larger than that of the central
engine due to the jet collimation into a narrower opening angle. The
observer at the intermediate angle $\theta_o=8^\circ$ initially
receives radiation mainly from the cocoon material. Its light curve
rises sharply, analogously to the light curves of observers at wider
angles. {Note that, despite the fact that the cocoon is less collimated 
than the jet, its luminosity can instantaneously exceed the input one 
due to the fact that the cocoon carries energy {\em stored} within the star
as the jet propagates. This stored energy gets released and converted into radiation
after the jet breaks out.}
The bright phase lasts for about 400~s, after which the
luminosity decreases. When the jet opening angle regains the injection
value at $t\sim3000$~s, however, the light curve seen by the
$\theta_o=8^\circ$ observer becomes very similar to the ones of  
observers with $\theta_o<5^\circ$. At even wider angles
$\theta_o>16^\circ$, observers only receive emission from the
cocoon, and the light curves decrease fairly quickly in a few hundred
seconds.

\section{Discussion}

We have performed an end-to-end simulation of an ULGRB from the
collapse of a BSG star.  Our stellar progenitor is a particularly
low-metallicity, rapidly-rotating massive star model. At core collapse,
such model fails to produce a canonical LGRB because of the presence
of a relatively extended envelope which sets an accretion timescale
much longer than that typical of LGRBs, and at the same time prevents
the very inner regions from retaining enough angular momentum to
satisfy the collapsar scenario. This model, while staying largely
mixed for the duration of its main sequence, does not evolve
quasi-chemically homogenously (Maeder 1987; Yoon \& Langer, 2005;
Woosley \& Heger 2006). It manages to build a compositional gradient,
which stops further mixing and allows to develop a core-envelope
structure. Due to the low-metallicity and the high helium fraction of
the envelope, this model only expands to few tens of solar radii
(compared to the $\approx 1000~\Rsun$ of typical RSGs), with the outer
layers unable to lose too much angular momentum through stellar
winds. At core collapse, such BSG model can produce an accretion disk
around a newly formed central object with accretion timescale
$\approx 10^4$ s. It is important to note that this particular solution
exists for a small region of the parameter space of initial rotational
velocity. That is, fixing all other parameters, models initialized
with slightly different values of initial rotational velocity can end
up producing either a WR (mostly for faster rotation) or a RSG (mostly
for slower rotation). A precise quantitative prediction of which stars
should end up producing BSG through this channel is difficult, since
the physics of internal chemical mixing depends on a number of poorly
understood processes. The important point is that if LGRBs are
produced through quasi chemically-homogenous evolution, then this
channel has to exist. This is because it operates in models with
similar masses and metallicities, but just rotating a bit slower than
those producing LGRBs. Therefore, if this channel dominates the
production of ULGRBs, these explosions should be found in environments
not too dissimilar from those of LGRBs (tracing similar stellar
populations). ULGRBs should also have lower rates than LGRBs, since
overall they require initial rotation rates in a smaller region of the
parameter space. These expected trends are only tentative: more
quantitative predictions require calculating grids of massive star
models of different mass, metallicity and with fine resolution in the
initial rotation rate, to be weighted with appropriate initial
functions for the various stellar parameters.

The final properties of our ULGRB model are fairly similar to those of
single BSG models studied by Woosley \& Heger (2012) in the context of
``Type 3 collapsars'', core collapse events for which only the surface
layers of the star have enough rotation to form a disk and the SN
shock is so weak that these layers are not ejected. Their BSG models
were calculated with initial metallicity $Z=1/10~Z_\odot$ and lower
initial rotation rate (about 20\% of the Keplerian value), showing
that similar outcomes can be produced for a less restrictive set of
initial parameters. In the specific case studied here, the choice of a
much lower metallicity and larger rotation rate was dictated by the
requirement to produce relatively compact and He-rich BSGs at core
collapse (our BSG model is more compact and He-rich than the one of
Woosley \& Heger 2012, with an envelope having an helium mass fraction larger than 0.7). 
A large surface helium abundance  is suggested by the observations of
ULGRB 11209A, for which the underlying superluminous supernova SN
2011kl was found to lack hydrogen lines in the spectra (Greiner et
al. 2015; but see also Ioka et al. 2016).

Finally, we remark that the scenario discussed here is not the only
possible way of producing rapidly rotating BSGs. These stars can be
a result of, e.g., binary stellar evolution and/or different
combinations of initial values of rotation, mass, metallicity as well
as choices of the physics of internal mixing (see e.g. Woosley \& Heger
2012). Because of this, we do not claim that the particular scenario
discussed in the paper should be favored. We just show that it is
possible to produce evolutionary calculations of BSGs to core collapse
that, as part of an end-to-end calculation, can produce the observed
properties of ULGRBs.

\acknowledgments RP acknowledges support from the NSF under grant
AST-1616157. DL acknowledges support from NASA grant NNX17AK42G. The
Flatiron Institute is supported by the Simons Foundation.

\section*{Software}
Python available from \url{python.org}, 
Matplotlib (Hunter 2007), ipython/jupyter (P\'erez \& Granger 2007, Kluyver et al. 2016), yt (Turk et al. 2011).  
\MESA\ inlists, data and the source code to produce the figures in this paper are available at \url{https://github.com/matteocantiello/ulgrb}

\end{document}